# High-NOON States With High Flux Of Photons Using Coherent Beam Stimulated Non-Collinear Parametric Down Conversion

Aziz Kolkiran,[1]

[1] Department of Engineering Sciences, İzmir Katip Çelebi University, Izmir 35620, Turkey.

Correspondence should be addressed to Aziz Kolkiran; aziz.kolkiran@ikcu.edu.tr

## Abstract

We show how to reach high fidelity *NOON* states with a high count rate inside optical interferometers. Previously, it has been shown that by mixing squeezed and coherent light at a beam splitter, it is possible to generate *NOON* states of arbitrary *N* with a fidelity as high as 94%. The scheme is based on higher order interference between "quantum" down-converted light and "classical" coherent light. However, this requires optimizing the amplitude ratio of classical to quantum light thereby limiting the overall count rate for the interferometric super-resolution signal. We propose using coherent-beam-stimulated non-collinear two-mode down converted light as input to the interferometer. Our scheme is based on the stimulation of non-collinear parametric down conversion by coherent light sources. We get a better flexibility of choosing the amplitude ratio in generating *NOON* states. This enables super-resolution intensity exceeding the previous scheme by many orders of magnitude. Therefore we hope to improve the magnitude of *N*-fold super-resolution in quantum interferometry for arbitrary *N* by using bright light sources. We give improved results for *N*=4 and 5.

## Introduction

Parametric down conversion (PDC) is a process that is used to produce light possessing strong quantum features. Photon pairs generated by this process show entanglement with respect to different physical attributes such as time of arrival [1] and states of polarization [2]. They are increasingly being utilized for very basic experiments to test the foundation of quantum mechanics and quantum information processing [2-4]. It is also recognized that entangled photon pairs could be useful in many practical applications in precision metrology involving e.g. interferometry [5, 7, 8, 18-20, 22], imaging [9, 10], lithography [11-14] , spectroscopy [15] and magnetometry[16]. There is a proposal [17] to use electromagnetic fields in *NOON* states to improve the sensitivity of measurements by a factor of *N*. In terms of photon number states, a two-mode field can be written as a superposition of two maximally distinguishable *N*-photon states

$$|NOON\rangle = \frac{1}{\sqrt{2}}(|N0\rangle + |0N\rangle). \quad (1)$$

Some implementations of this state exist [18-20]. The *N*-photon coherence is optimally sensitive to small phase shifts between the two modes. In particular, the use of photon pairs in interferometers allows phases to be measured to the precision in the Heisenberg limit where uncertainty scales as $1/N$ [19] as compared to the shot noise limit where it scales as $1/\sqrt{N}$. This means that for large number of particles, a dramatic improvement in measurement resolution should be possible. There are various methods for generating path entangled states with high photon numbers up to *N=5* [18-20]. Afek et. al. [22] experimentally realized high fidelity *NOON* states for *N=2,3,4* and *5* in a single setup. They realized the idea introduced by Hoffmann and Ono[23] by mixing coherent state and squeezed vacuum or degenerate parametric down conversion (PDC) (both at single modes) at a 50/50 beam-splitter (BS) in two input ports. The idea is based on the indistinguishability of the photons after a BS. The indistinguishable processes always result in quantum interferences. This means that there is enhancement or suppression of some terms in the superposition. The setup for the observation of path entanglement in the interference of coherent state light and downconverted photon pairs is shown in Figure 1(a). The two beams interfere at the input beam splitter. This method can be interpreted as a cancellation of all output components other than $|N0\rangle$ and $|0N\rangle$ by destructive quantum interference. As an example, Figure 1(b,c) shows that there are 2 ways to combine 3 photons at the BS from the sources of coherent light and PDC. These 2 ways mixes the photons coming from coherent light and PDC. The exact cancellation of unwanted terms cannot be achieved at photon numbers higher than 3 [24], so that the extension to higher photon numbers appears to be difficult. Nevertheless, one can get close to exact path entangled *N*-photon states with the fidelities of 92% or more for arbitrary *N*. Particularly, it has been shown in [23] that the fidelity of 94.1% can be reached for *N=5* and in [22], Afek et.al demonstrated this experimentally with visibility of 42%. The *NOON* state fidelity of the output state can be optimized by tuning the relative strength of average photon numbers between the coherent light and PDC. It appears that the determination of this relative strength depends on the number of ways mixing the two sources. The best results were achieved for the low gain limit of the PDC source. There is also a proposal [25] to increase the generation probability of *NOON* states by mixing even/odd coherent states.

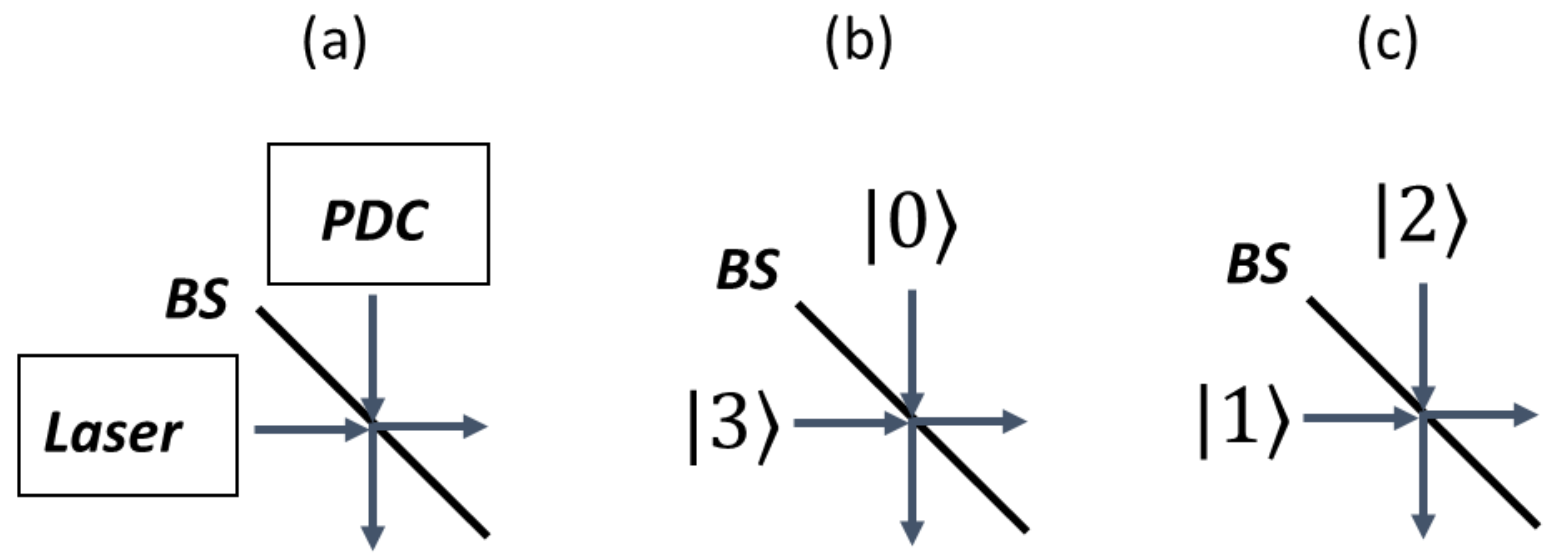

Figure 1: (a) The setup for mixing laser and PDC lights at a BS. There are 2 distinct ways to mix laser and PDC photons leading to 3-NOON state: (b) No photons from PDC and 3 photons from laser, (c) 2 photons from PDC and 1 photon from laser.

In this work, we propose a new idea [26] using stimulated parametric processes along with spontaneous ones to produce path entangled states of arbitrarily high photon number *N* with fidelity greater than 90% at strong gain regime meaning that the sources are bright. The stimulated processes enhance the count rate by several orders of magnitude. We use coherent beams at the signal and idler frequencies of a non-collinear PDC source (see Fig. 2). We further find that the phases of coherent fields can also be used as tuning knobs to control both the fidelity and the magnitude (repetition rate) of *NOON* state intensity. It may be borne in mind that the process of non-collinear spontaneous parametric down conversion has been a workhorse for more than two decades in understanding a variety of issues in quantum physics and in applications in the field of imaging and sensing.

## Materials and Methods

We now describe the idea and the results of preliminary calculations that support the above assertion. Consider the scheme shown in Figure 2. Here $\hat{a}_1$ and $\hat{b}_1$ are the signal and idler modes driven by the coherent fields. The usual case of spontaneous parametric down conversion is recovered by setting $\alpha_0 = \beta_0 = 0$. The $\psi$ is the phase introduced by the object or by an interferometer. For down conversion of type II the signal and idler would be two photons in two different states of polarization.

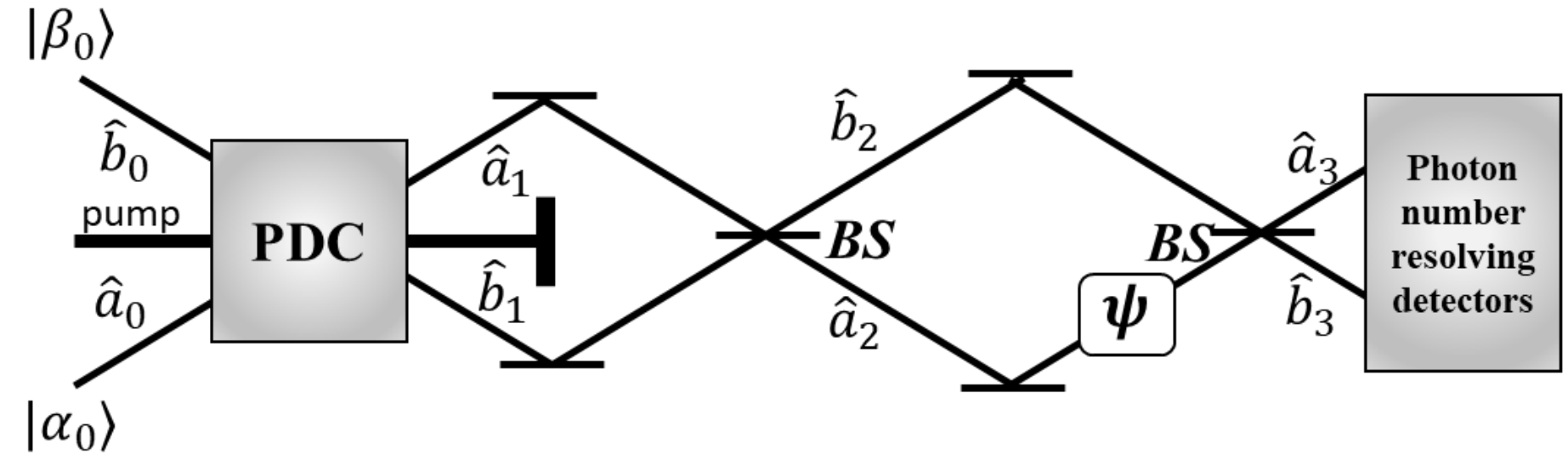

Figure 2: The setup for the quantum interferometer using coherent beam stimulated PDC. The modes $\hat{a}_1$ and $\hat{b}_1$ are driven by coherent beams $\alpha_0, \beta_0$ and non-collinear PDC modes. They enter the Mach-Zehnder interferometer through the first beam splitter (BS) and at the exit BS they are detected in a coincidence measurement by the photon number resolving detectors.

The input state before the first beam-splitter is defined by

$$|\alpha_0, \beta_0\rangle_{(r,\phi)} = D(\alpha_0, \beta_0) S(r, \phi)|0\rangle, \quad (2)$$

where $|0\rangle$ is the two-mode vacuum state. The two-mode down conversion operator,

$$S(r, \phi) = exp[r(a_0 b_0 e^{-i\phi} - a_0^\dagger b_0^\dagger e^{i\phi})], \quad (3)$$

is characterized by a real gain parameter $r$, and phase, $\phi$, that determines the phase of the down converted photons. The two-mode displacement operator

$$D(\alpha_0, \beta_0) \equiv D_{a_0}(\alpha_0) D_{b_0}(\beta_0), \quad (4)$$

is a product of displacement operators for each mode. The input state in the Schrödinger picture can be written as the following superposition,

$$|\alpha_0, \beta_0\rangle_{(r,\phi)} = \sum_{N=0}^{\infty} |\Psi_N\rangle, \quad (5)$$

where $|\Psi_N\rangle$ is the *N*-photon component given by,

$$|\Psi_N\rangle = \sum_{m=0}^{N} C(m, N-m)|m, N-m\rangle, \quad (6)$$

where $C(m, N-m)$ are the coefficients of N-photon states in the input state $|\alpha_0, \beta_0\rangle_{(r,\phi)}$ of the interferometer shown in Figure 2. For $\alpha_0 = \beta_0 = 0$, the coefficients are given by $C(m,n) = \delta_{mn} \frac{\left(-e^{i\phi} tanh(r)\right)^n}{cosh(r)}$. It is easy to produce two-photon *NOON* state by using Hong-Ou-Mandel scheme [27]. However for *N*>*2*, we need some extra parameters to cancel the unwanted terms inside the interferometer. This can be done by using coherent states as seeds, initially in the vacuum modes of the non-collinear down conversion (see Figure 2). When the stimulation is on, the coefficients $C(m, n)$ become functions of coherent field amplitude and it can be expressed in a closed form [28],

$$C(m,n) = \frac{(-tanh(r))^p}{cosh(r)} \left(\frac{p!}{q!}\right)^{1/2} \mu_1^{m-p} \mu_2^{n-p} L_p^{(q-p)} \left(\frac{\mu_1 \mu_2}{tanh(r)}\right) e^{-(\alpha_0^* \mu_1 + \beta_0^* \mu_2)/2} \quad (7)$$

where

$$p = min(m, n) \quad (8)$$

$$q = max(m, n) \quad (9)$$

$$\mu_1 = \alpha_0 + \beta_0^* \, tanh(r) \quad (10)$$

$$\mu_2 = \beta_0 + \alpha_0^* \, tanh(r) \quad (11)$$

We take $\alpha_0 = \beta_0 = |\alpha_0| e^{i\theta}$ and $\phi = 0$ for simplicity. We use the phase of coherent fields, $\theta$, for controlling the interference between coherent states and down converted photons. This interference results from the indistinguishability between coherent state photons and down converted photons [26] because their polarization, frequency and momentum states are completely the same. The creation of an ideal *NOON* state would require suppression of all the non-*NOON* components after the first beam splitter. By tuning the parameters available in the scheme, we can reach this with a very high fidelity using multiphoton interference. The fidelity of the output state's normalized *N* photon component with a *NOON* state is

$$F_N = |\langle NOON|U_{BS}|\Psi_N^{norm}\rangle|^2, \quad (12)$$

where $U_{BS}$is the 50:50 beam-splitter unitary transformation and $|\Psi_N^{norm}\rangle$ is given by Equation (6) with a normalization constant. There are more than one way to represent $U_{BS}$ and here we adapt the choice given in [29]. It is easy to calculate the fidelity given by (12) by the inner product of the states $U_{BS}|NOON\rangle$ and $|\Psi_N^{norm}\rangle$. The beam-splitter transformation of the *NOON* state is

$$U_{BS}|NOON\rangle = \frac{1}{\sqrt{2^{N-1}}} \sum_{m=0}^{N} \sqrt{\binom{N}{m}} \cos\left\{\frac{\pi}{2}\left(m - \frac{N}{2}\right)\right\} |m, N-m\rangle. \quad (13)$$

As an example let us consider the calculation of $F_4$. The beam-splitter transforms $|4004\rangle$ to be,

$$U_{BS}|4OO4\rangle = -\frac{1}{2\sqrt{2}}(|40\rangle + |04\rangle) + \sqrt{\frac{3}{4}}|22\rangle, \quad (14)$$

and Equations (6)-(14) would lead to

$$F_4 = |-\frac{1}{2\sqrt{2}}\left(C(4{,}0) + C(0{,}4)\right) + \sqrt{\frac{3}{4}}C(2{,}2)|^2 \quad (15)$$

upto a normalization. The interferometric phase measurement is done by the photon number resolving detection (see Figure 2). For example, for four-fold resolution enhancement, we use an array of four single photon coincidence counting modules in 2 by 2 arrangement as proposed in [16] (it means two of detectors are in the upper exit port and two of them in the lower exit port) or 3 by 1 scheme as proposed in [30].

There are two parameters to optimize the fidelity once we fix the phase and the flux of PDC photons; the phase of the coherent fields, $\theta$, and the pair amplitude ratio of the coherent state and PDC state which is given by

$$\gamma = |\alpha_0|^2/r \quad (16)$$

for weak fields. For the strong fields this ratio should be replaced by

$$\gamma = |\alpha_0|^2/sinh^2(r) \quad (17)$$

because the magnitude of pair flux from PDC is dominated by the term $sinh^2(r)$ [26]. This optimization shows a different character for weak and strong field regimes. In the weak field regime the maximum fidelity is very sensitive to $\gamma$ and it is a limiting factor for the total strength of the total signal. On the other hand, in the limit of high gain, $\gamma$ takes larger values together with flexibility in the optimization. This makes possible the super-resolving phase measurements with high *NOON* states at much brighter light resources.

We would like to explain the physics behind the advantage of using our scheme compared to the idea of Hoffmann and Ono [23]. Suppose that we want to construct *N=3 NOON* state. In our scheme, there are six distinct ways to mix three photons coming from the laser light (coherent states) and PDC (see Fig. 3). This is three times larger than the scheme used in [22,23]. For *N=4* and *N=5*, in our scheme there are 9 and 12 ways respectively while there are 2 and 3 ways in [22,23]. The number of photon mixing ways is getting even larger and larger for higher *NOON* states. As a result, our scheme have higher order interference effects leading to higher count rates with similar fidelity. We give detailed results in the next section.

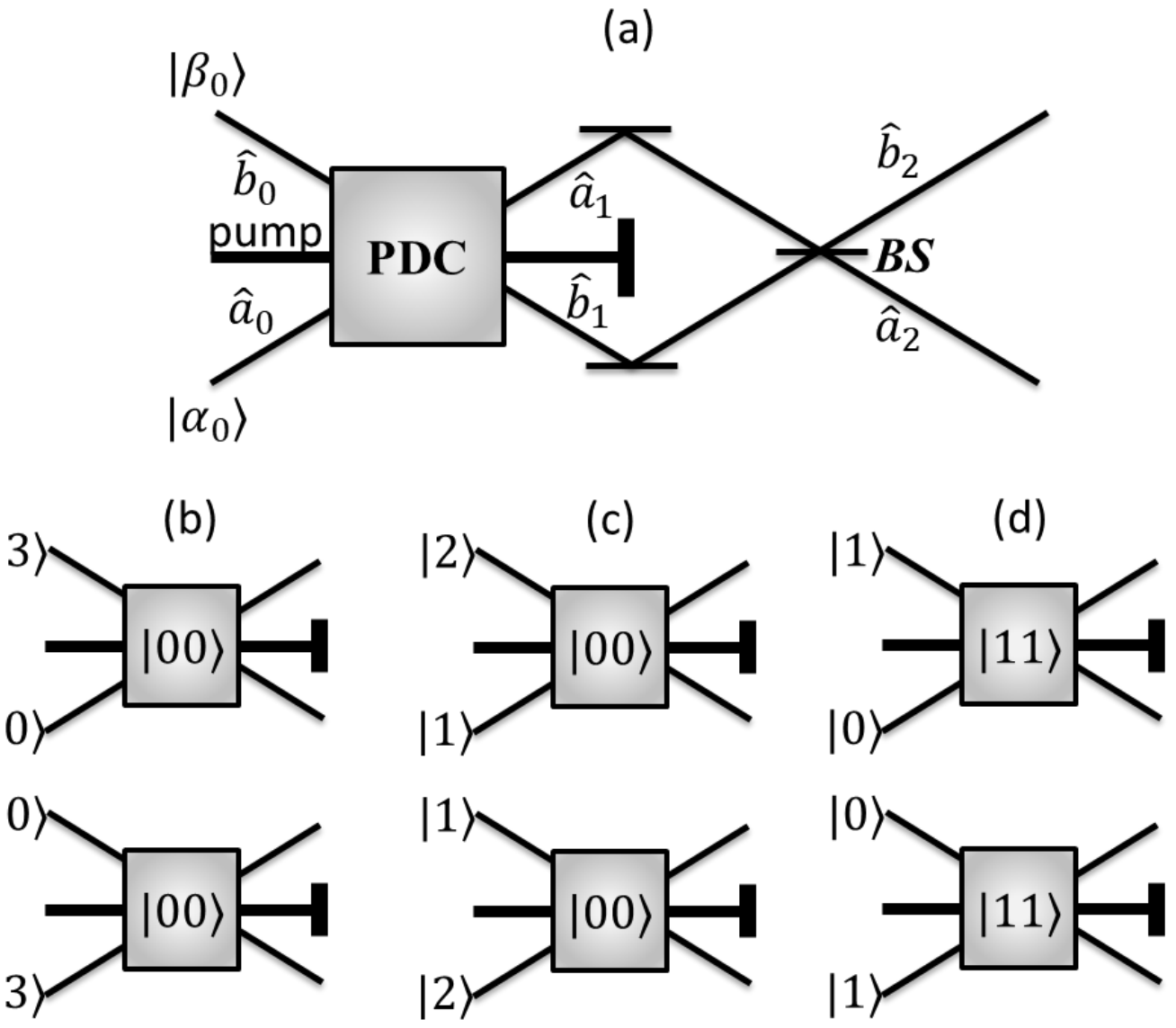

Figure 3: (a) The setup for mixing laser and non-collinear PDC lights at a BS. There are six distinct ways to mix laser and PDC photons leading to 3-NOON state: (b) No photons from PDC, 3 photons from upper laser, no photons from lower laser and vice versa. (c) No photons from PDC, 2 photons from upper laser, 1 photon from lower laser and vice versa. (d) 2 photons from PDC, 1 photon from upper laser , no photons from lower laser and vice versa. Compare this to Figure 1 (b,c).

## Results and Discussion

We now give our results of the theoretical fidelity for the generated *NOON* states in Figure 4 and Figure 5. Figure 4(a) shows the fidelity of interferometer state's N photon component with *NOON* state with *N=4* for weak fields. The horizontal axis is the phase of the coherent field. The phase of PDC is chosen to be zero. Here the gain parameter of the PDC is $r = 0.1$ and the amplitude ratio of the coherent state and PDC given by $\gamma = |\alpha|^2/r$ is optimized at 2.26 for the maximum fidelity of $F_4 = 93.3\%$. The maximum fidelity is reached at phases of 0 and π. This is much better than the case in which $\alpha = 0$ (with down converted photons only) with a fidelity of 75% and the case in which $r = 0$ (coherent fields only) with a fidelity of 50%. Figure 4(b) shows the fidelity profile in the limit of high gain with $r = 4.5$ (this gain value has been reported experimentally in [31]) . The black, red and blue curves are for $\gamma = 10, 50$ *and* $150$ with fidelities of 92%, 90% and 81% respectively. It is clear from the plots that we have much more flexibility in the amplitude ratio. By choosing a larger amplitude ratio we can reach a total flux of coherent state photons having approximately two orders of magnitude higher than PDC photons. For example, for $\gamma = 50$, when the mean number of photons in PDC $sinh^2(4.5) \approx 2 \times 10^3$ we have $\approx 10^5$ coherent state photons. In physical terms, as proposed by [20], $\gamma^2$ is the two-photon probability of the classical source divided by that of the quantum source. In this case we have $\gamma^2 = 2500$, meaning that for every pair of PDC photons, 2500 pairs of laser photons are used. Although the fidelity is down only by 3% as proposed in [22] with $\gamma^2 = 3$, we reach three orders of magnitude in signal repetition rate. This would increase the coincidence count rate tremendously when the method is used in experimental quantum interferometry. We note here that a slight decrease in the fidelity would only decrease the visibility slightly. As noted in [23], the phase sensitivity of an *N*-photon fringe with visibility *V* is given by $\delta\psi = 1/VN$, we can expect a phase sensitivity of *1/0.90N* in the high photon number limit, a result that is only slightly lower than the Heisenberg limit of *1/N*, achieved by maximal path entanglement. In the strong field regime, one can notice the strange dependence of fidelity on the coherent state phase $\theta$. This is unlike the weak-field regime, which shows a nice periodic dependence. In the strong-field regime, as $\gamma$ gets larger, the bandwidth of phase dependence for the maximum fidelity gets narrower in some sense. The physics of this behaviour needs to be investigated further.
Figure 5 shows the results for *N=5*. Figure 5(a) shows the optimized fidelity (maximum at 91%) profile at the low gain limit, $r = 0.1$ and $\gamma = 0.6$ . Figure 5(b) shows the fidelity profile in the limit of high gain with $r = 4.5$. The black, red and blue curves are the fidelity profiles for $\gamma = 10, 50\ and\ 150$ respectively. The respective maximum fidelities are approximately 91%, 88% and 84%. Here, the fidelity of 91% is being reached when $\gamma^2 = 100$ meaning that for every pair of PDC photons, 100 pairs of laser photons are used to see five-photon interference pattern.

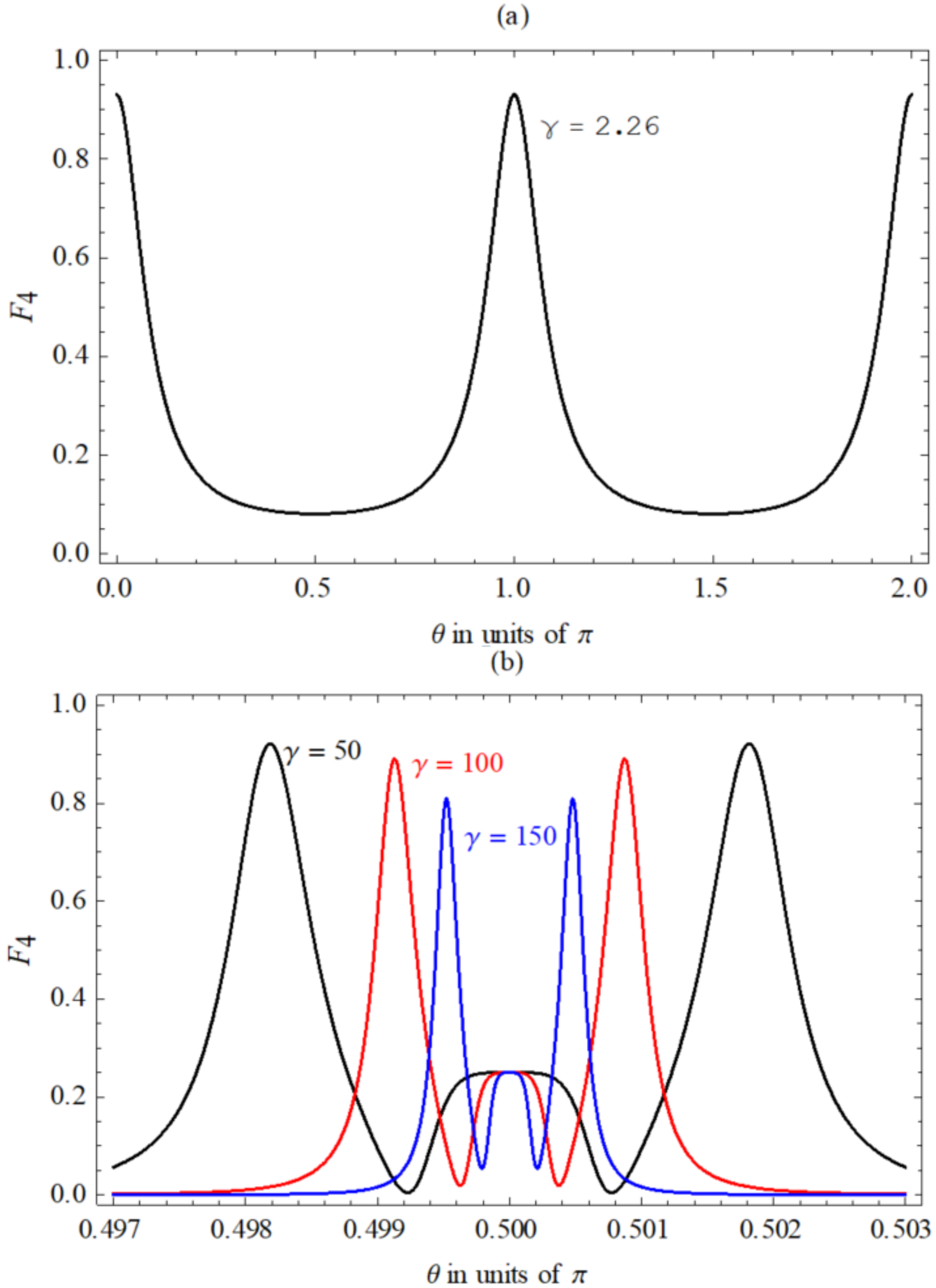

Figure 4: The fidelity plots for *4-NOON* states. (a) The fidelity of interferometer state's *N* photon component with *NOON* state with *N=4* for weak fields. The horizontal axis is the phase of the coherent field. The phase of PDC is chosen to be zero. Here the gain parameter of the PDC is $r = 0.1$ and the amplitude ratio is optimized at $\gamma = |\alpha|^2/r = 2.26$ for the maximum fidelity of $F_4 = 93.3\%$. The maximum fidelity is reached at phases of *0* and $\pi$. This is much better than the case in which (with down converted photons only) with a fidelity of 75% and the case in which (coherent fields only) with a fidelity of 50%. (b) The fidelity profile in the limit of high gain with $r = 4.5$ (this gain has been reported in [31]) . The black, red and blue curves are with fidelities of 92%, 90% and 81% respectively. It is clear from the plots that we have much more flexibility in pair amplitude ratio. By choosing a larger amplitude ratio we can reach a total flux of coherent state pair photons having approximately 3 orders of magnitude higher than PDC pair photons.

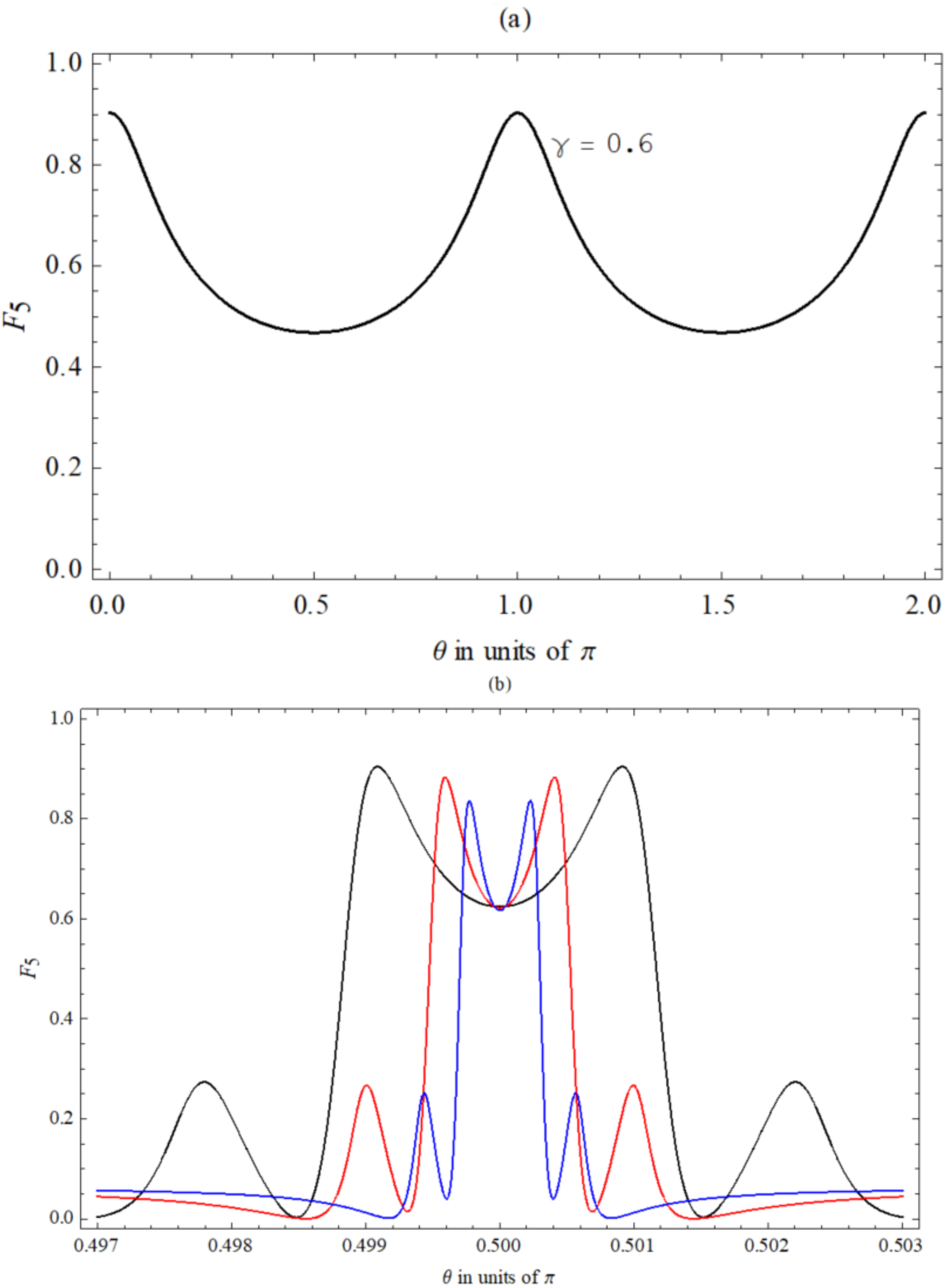

Figure 5: Same with fig.2 for *N=5*; (a) The optimized fidelity (maximum at 91%) profile at the low gain limit, $r = 0.1$ and (b) The fidelity profile in the limit of high gain with $r = 4.5$. The black, red and blue curves are the fidelity profiles for $\gamma = 50, 100\ and\ 150$ respectively. The respective maximum fidelities are approximately 91%, 88% and 84%.

## Conclusions

In conclusion, we have shown that using stimulating coherent fields in the non-collinear PDC setup generates not only high-*NOON* states with large *N* but also with arbitrary intensity for *N=4* and *5*. The realistic application of *NOON* states in quantum metrology requires high intensity flux of photons. The theoretical improvement by using coherent field stimulated non-collinear PDC photons over the method of mixing squeezed light with coherent state [22,23] implies a fundamental connection between non-locality of the source and creation of *NOON* states. The number of ways mixing non-classical photons with the classical ones in the proposed scheme is multiple times larger which leads to higher order interference effects. This is why the scheme is using more classical photons over non-classical ones (three orders of magnitude larger) to reach Heisenberg limited sensitivity. The ongoing development of high gain parametric down conversion together with efficient detectors shows promise for realizing the scheme proposed in this paper. We also state that the scheme can be applied to utilize *NOON* states of higher *N* number. Further study to this end is required.

## Data Availability

The analytic data used to support the findings of this study are included within the article. The plots in Figure 2 and Figure 3 can be obtained by plotting the closed forms given by Equations (6)-(15) using an appropriate “function plotting software”, such as mathematica or matlab.

## Conflicts of Interest

The author declares that there is no conflict of interests regarding the publication of this paper.

## Funding Statement

This research is supported by the Izmir Katip Celebi University, Coordinatorship of Scientific Research Projects under grant 2013-2-FMBP-34.